\documentclass{ws-procs975x65}
\usepackage{color}
 \usepackage{hyperref}
 \hypersetup{colorlinks=true, urlcolor 	=blue}
\usepackage{hyperref}

\begin{document}

\title{\uppercase{Transport techniques for non-Gaussianity}}

\author{\uppercase{David J. Mulryne }}
%$^*$}}

\address{School of Physics and Astronomy, Queen Mary University of London,\\
London, E1 4NS, United Kingdom}
%$^*$E-mail: D.Mulryne@qmul.ac.uk}

\begin{abstract}
This 
proceedings 
contribution provides 
a brief update on  
the transport approach to calculating 
the statistics of perturbations produced during inflation. It is based in particular on work in 
collaboration with Gemma Anderson and David Seery\cite{Anderson:2012em}.

\end{abstract}


\bodymatter

\section{Inflationary statistics}\label{sect:inflation}
During inflation, quantum fluctuations 
become classical perturbations when their wavelengths becomes larger 
than the cosmological horizon. Different models of inflation produce  
perturbations with different statistical properties, and we wish to track 
the evolution of these statistics. 
Transport techniques\cite{Anderson:2012em, Mulryne:2009kh, Mulryne:2010rp} 
are one method to do this. They offer analytic insights into the evolution\cite{Elliston:2011dr, Seery:2012vj}, and 
reduce the problem to a set of coupled ordinary differential equations 
that can easily be solved numerically.

For most applications we are interested in just the first 
few cummulants (inflationary correlation functions) of the Fourier 
modes of the curvature perturbation $\zeta$.
These are related in a straightforward manner to the statistics 
of the field perturbations, which in 
multi-field canonical models 
are very close to Gaussian at horizon crossing\cite{Seery:2005gb} but evolve thereafter. In these proceedings, we 
review how this super-horizon evolution of statistics can be calculated. 

\section{Evolution of field space statistics and `shapes'}

We begin with the evolution equation for the Fourier modes of 
the field fluctuations themselves. These can be written in a DeWitt index form\cite{Seery:2012vj, Anderson:2012em} in which
a compound primed index includes a field label $\alpha$ and a momentum label
$\mathbf{k}_\alpha$. 
The summation convention applied to $\alpha'$
implies integration over momentum with measure
${\rm d}^3 \mathbf{k}_\alpha / (2\pi)^3$, and
summation over the field species.
The equations are
\begin{eqnarray}
		\frac{{\rm d} \delta \varphi_{\alpha'}}{{\rm d} N}
		= 
			u_{\alpha' \beta'} \delta \varphi_{\beta'}
			&+&
			\frac{1}{2!}
			u_{\alpha' \beta' \gamma'}
			\Big(
				\delta \varphi_{\beta'} \delta \varphi_{\gamma'}
				-
				\langle
					\delta \varphi_{\beta'} \delta \varphi_{\gamma'}
				\rangle
			\Big)
			\nonumber \\&+&
			\frac{1}{3!}
			u_{\alpha' \beta' \gamma' \delta'}
			\Big(
				\delta \varphi_{\beta'} \delta \varphi_{\gamma'}
				\delta \varphi_{\delta'}
				-
				\langle
					\delta \varphi_{\beta'} \delta \varphi_{\gamma'}
					\delta \varphi_{\delta'}
				\rangle
			\Big)
			+
			\cdots\,,
	\label{eq:jacobi}
\end{eqnarray}
where the time variable $N$ is e-folds. The $u$ coefficients take the form $u_{\alpha^{\prime}\beta^{\prime}} 
	\equiv		(2\pi)^3
		\delta(\mathbf{k}_{\alpha} - \mathbf{k}_{\beta})
		u_{\alpha\beta}[N]$,
	$u_{\alpha^{\prime}\beta^{\prime}\gamma^{\prime}} 
	\equiv
		(2\pi)^{3}
		\delta(\mathbf{k}_{\alpha} - \mathbf{k}_{\beta} - \mathbf{k}_{\gamma})
		u_{\alpha\beta\gamma}[N]$ and
	$u_{\alpha^{\prime}\beta^{\prime}\gamma^{\prime}\delta^{\prime}} 
	\equiv
		(2\pi)^{3}
		\delta(\mathbf{k}_{\alpha} - \mathbf{k}_{\beta} - \mathbf{k}_{\gamma}
		- \mathbf{k}_{\delta})
		u_{\alpha\beta\gamma\delta}[N]$, where $u$ with unprimed indices is 
just a function of unperturbed background quantities, and hence of $N$. Here we have assumed slow-roll and so neglected field velocity perturbations (this can be 
easily relaxed).

Utilising the simple principal ${\rm d }\langle A \rangle / {\rm d} N= \langle  {\rm d} A / {\rm d } N \rangle$, which follows 
from probability conservation\cite{Seery:2012vj}, and identifying $A$ with products of field perturbations, we find 
\begin{eqnarray}
	\frac{{\rm d} \Sigma_{\alpha^{\prime}\beta^{\prime}}}{ {\rm d} N}
	&=&
		u_{\alpha^{\prime}\gamma^{\prime}}
		\Sigma_{\gamma^{\prime}\beta^{\prime}}
		+
		u_{\beta^{\prime}\gamma^{\prime}}
		\Sigma_{\gamma^{\prime}\alpha^{\prime}}
		+\dots ,\nonumber \\
\frac{{\rm d} \alpha_{\alpha^{\prime}\beta^{\prime}\gamma^{\prime}}}{{\rm d} N}
	&=&
		u_{\alpha^{\prime}\lambda^{\prime}}
		\alpha_{\lambda^{\prime}\beta^{\prime}\gamma^{\prime}}
		+
		\left (u_{\alpha^{\prime}\lambda^{\prime}\mu^{\prime}}
		\Sigma_{\lambda^{\prime}\beta^{\prime}}
		\Sigma_{\mu^{\prime}\gamma^{\prime}}
		+
		\text{3 cyclic}\right)+\dots ,\nonumber \\
\frac{{\rm d} \beta_{\alpha' \beta' \gamma' \delta'}}{{\rm d} N}
	&=& 
		\Big(
			u_{\alpha' \lambda'}
			\beta_{\lambda' \beta' \gamma' \delta'}
			+
			\text{3 cyclic}
		\Big)+
			\Big(
			u_{\alpha' \lambda' \mu'}
			\alpha_{\lambda' \beta' \gamma'} \Sigma_{\mu' \delta'}
			+
			\text{11 cyclic}
		\Big) 	\nonumber \\&+&
		\Big(
			u_{\alpha' \lambda' \mu' \nu'}
			\Sigma_{\lambda' \beta'} \Sigma_{\mu'\gamma'} \Sigma_{\nu' \delta'}
			+
			\text{3 cyclic}
		\Big)
		+ \dots ,
\end{eqnarray}
where the dots indicate each equation is truncated at leading order, and $\Sigma_{\alpha' \beta'}=\langle \delta \phi_{\alpha'} \delta \phi_{\beta'} \rangle$, $\alpha_{\alpha' \beta' \gamma'} = \langle \delta \phi_{\alpha'} \delta \phi_{\beta'} \delta \phi_{\gamma'} \rangle$ and $\beta_{\alpha' \beta' \gamma' \delta'} = 	\langle
			\delta \varphi_{\alpha^{\prime}}
			\delta \varphi_{\beta^{\prime}}
			\delta \varphi_{\gamma^{\prime}}
			\delta \varphi_{\delta^{\prime}}
		\rangle
		-
		\Sigma_{\alpha' \beta'} \Sigma_{\gamma' \delta'}
		-
		\Sigma_{\alpha' \gamma'} \Sigma_{\beta' \delta'}
		-
		\Sigma_{\alpha' \delta'} \Sigma_{\beta' \gamma'} $. These represent the equations of motion 
for cummulants/correlations of the field fluctuations.

On super-horizon scales, not all `shapes' of correlations are generated. Starting with 
Gaussian fluctuations with close to scale invariant 
power spectrum, one finds\cite{Anderson:2012em}
\begin{eqnarray}
\Sigma_{\alpha^{\prime}\beta^{\prime}}
	&=&
		(2\pi)^3
		\delta(\mathbf{k}_\alpha + \mathbf{k}_\beta)
		\frac{\Sigma_{\alpha\beta}}{k_{\alpha}^{3}}\,,\nonumber \\ 
	\alpha_{\alpha^{\prime}\beta^{\prime}\gamma^{\prime}}
&=&		(2\pi)^3\delta(\mathbf{k}_\alpha + \mathbf{k}_\beta + \mathbf{k}_\gamma)
		\left(
			\frac{a_{\alpha\mid\beta\gamma}}{k_{\beta}^{3}k_{\gamma}^{3}}
			+
			\frac{a_{\beta\mid\alpha\gamma}}{ k_{\alpha}^{3}k_{\gamma}^{3}}
			+
			\frac{a_{\gamma\mid\alpha\beta}}{k_{\alpha}^{3}k_{\beta}^{3}}
		\right)\,,\nonumber \\ 
	\beta_{\alpha^{\prime}\beta^{\prime}\gamma^{\prime}\delta^{\prime}}
	&=&
		(2\pi)^3
		\delta(\mathbf{k}_\alpha + \mathbf{k}_\beta + \mathbf{k}_\gamma
			+ \mathbf{k}_\delta)
		\Big(
			\frac{g_{\alpha\mid\beta\gamma\delta}}
				{k_\beta^3k_\gamma^3k_\delta^3}
			+ \text{3 cyclic}
\nonumber \\ &+& \frac{\tau_{\alpha\beta\mid\gamma\delta}}
				{k_\alpha^3k_\beta^3 |\mathbf{k}_\alpha + \mathbf{k}_\gamma|^3}
			+ \text{11 cyclic}
		\Big)\,,
\end{eqnarray} 
where $\Sigma_{\alpha \beta}$, $a_{\alpha\mid \beta \gamma}$, $g_{\alpha\mid\beta\gamma\delta}$ and $\tau_{\alpha\beta\mid\gamma\delta}$
carry only very mild scale dependence. These shape parameters obey their own transport equations\cite{Anderson:2012em}
\begin{eqnarray}
\frac{{\rm d} a_{\alpha\mid\beta\gamma}}{{\rm d} N}
	&=&
		u_{\alpha\lambda} a_{\lambda\mid\beta\gamma}
		+
		u_{\beta\lambda} a_{\alpha\mid\lambda\gamma}
		+
		u_{\gamma\lambda} a_{\alpha\mid\beta\lambda}
		+
		u_{\alpha\lambda\mu} \Sigma_{\lambda\beta} \Sigma_{\mu\gamma}\,, \nonumber \\
\frac{{\rm d} g_{\alpha\mid\beta\gamma\delta}}{{\rm d} N}
	& =&
		u_{\alpha\lambda} g_{\lambda\mid\beta\gamma\delta}
		+
		\big( u_{\beta\lambda} g_{\alpha\mid\lambda\gamma\delta}
		+
                     u_{\alpha\lambda\mu} a_{\lambda\mid\beta\gamma} \Sigma_{\mu\delta}	+ \text{3 cyclic} \big)
          +                     u_{\alpha\lambda\mu\nu}
			\Sigma_{\lambda\beta}
			\Sigma_{\mu\gamma}
			\Sigma_{\nu\delta}\,,\nonumber \\
\frac{{\rm d} \tau_{\alpha\beta\mid\gamma\delta}}{{\rm d} N}
	& =&
		u_{\alpha\lambda} \tau_{\lambda\beta\mid\gamma\delta}
		+
		u_{\beta\lambda} \tau_{\alpha\lambda\mid\gamma\delta}
		+
		u_{\gamma\lambda} \tau_{\alpha\beta\mid\lambda\delta}
		+
		u_{\delta\lambda} \tau_{\alpha\beta\mid\gamma\lambda}
		\nonumber \\&+&
		u_{\gamma\lambda\mu} \Sigma_{\mu\alpha} a_{\delta\mid\lambda\beta}
		+
		u_{\delta\lambda\mu} \Sigma_{\mu\beta} a_{\gamma\mid\lambda\alpha}\,,
\end{eqnarray}
and can used to form the correlation functions of $\zeta$, and the calculate the related non-Gaussian parameters of $\zeta$, namely   
 $f_{\rm NL}$, $\tau_{\rm NL}$ and $g_{\rm NL}$ (see Refs.~\refcite{Lyth:2005fi, Seery:2006js, Byrnes:2006vq}) 
as explained in detail in Ref.~\refcite{Anderson:2012em}. These equations, therefore, represent a practical algorithm to evolve  
$f_{\rm NL}$, $\tau_{\rm NL}$ and $g_{\rm NL}$ from horizon crossing.

\begin{figure}[t]
\begin{center}
\psfig{file=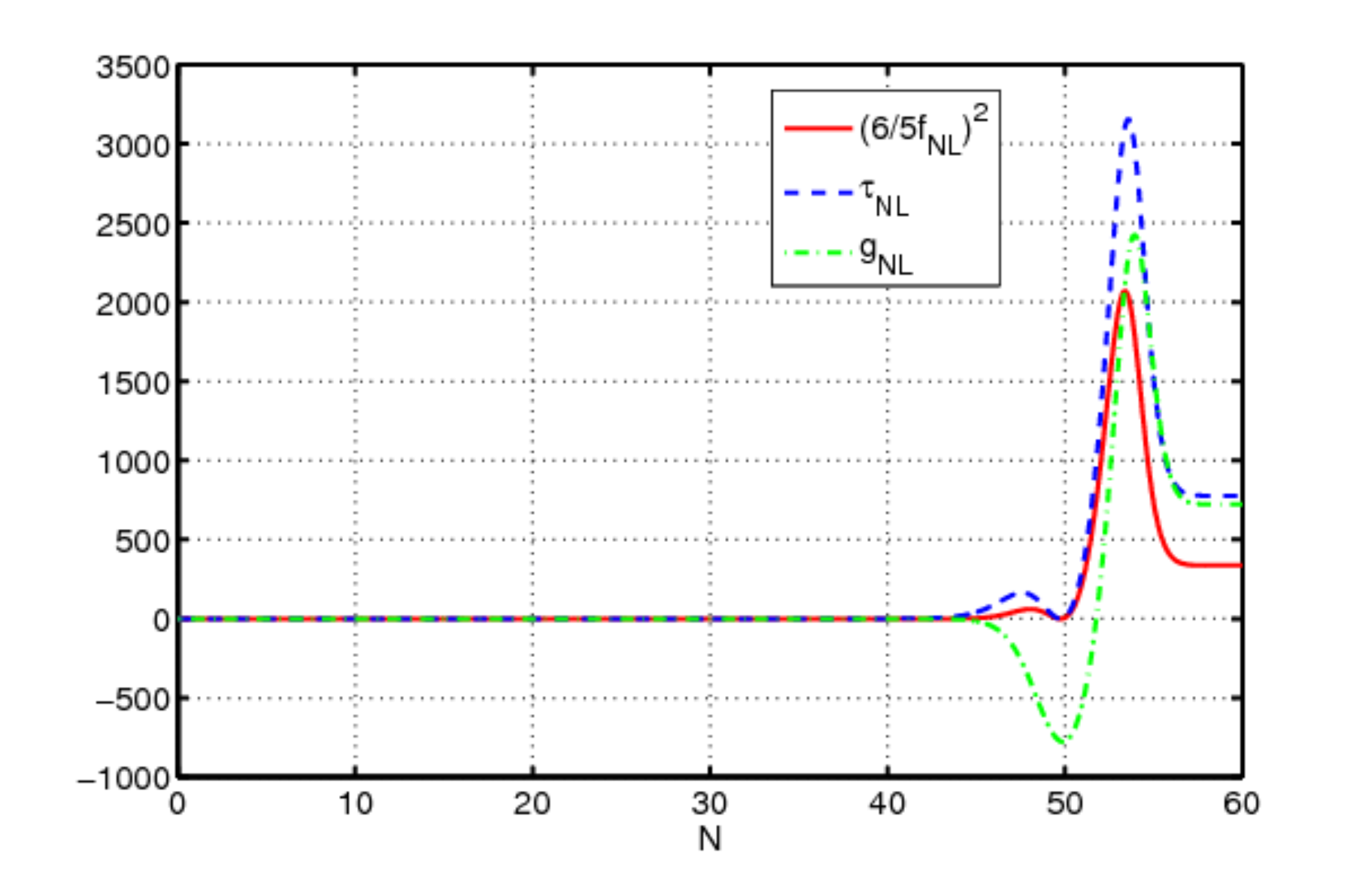,width=3.25in}
\end{center}
\caption{The evolution of  $f_{\rm NL}$, $\tau_{\rm NL}$ and $g_{\rm NL}$ from horizon-crossing to the end of inflation
for the model of inflation defined in the text.}
\label{fig1}
\end{figure}

For example, numerically solving the equations for the hybrid potential model with two light fields\cite{Mulryne:2011ni},
$V= M^4
		\left [
			\frac{1}{2} m^2 \phi_1^2
			+
			\frac{1}{2}g^2\phi_1^2 \phi_2^2
			+
			\frac{\lambda }{4} \left ( \phi_2^2-v^2 \right)^2
		\right ]  
$, with parameter choices and initial conditions $g^2=v^2/\phi_{{\rm crit}}^2$, $m^2=v^2$, $v=0.2 M_{\rm pl}$, $\phi_{{\rm crit}}=20 M_{\rm pl}$, $\lambda=5$, $\phi_{1{\rm exit}} = 15.5 M_{\rm pl}$ and $\phi_{2 {\rm exit}} = 0.0015 M_{\rm pl}$, results in the evolution of the non-Gaussianity parameters plotted in Fig.~\ref{fig1}.

\section{Discussion}

Space has only permitted a brief review, so it is worth mentioning work on the transport approach to inflationary observables has yielded a number of other important results. In particular, we glossed over how the statistics of field perturbations are 
converted to those of $\zeta$, and a particularly simple derivation of these relations can be found\cite{Anderson:2012em}. Moreover, the 
correlations involved in the transport method can be related to the coefficients of a $\delta N$\cite{Lyth:2005fi} style Taylor expansion, leading 
to transport equations for the 
Taylor coefficients themselves\cite{Seery:2012vj,Anderson:2012em}. Furthermore, a geometrical decomposition of the transport equations is possible\cite{Seery:2012vj}. Finally, current work is ongoing in extending the transport method to sub-horizon scales\cite{mulryne:2013uka} and providing reusable numerical tools based on transport methods.

\section*{Acknowledgements}
DJM is supported by Science and Technology Facilities Council
grant ST/J001546/1S. Thanks to my collaborators Gemma Anderson and David Seery.

\bibliographystyle{ws-procs975x65}
\bibliography{paper.bib}

\end{document}